# Predictive Simulations of ST40 Hot Ion Plasmas using the TGLF model

M.S. Anastopoulos Tzanis [1], M.R. Hardman [1], Y. Zhang [2], X. Zhang [1], A. Sladkomedova [1], A. Dnestrovskii [1], Y.S. Na [3], J.H. Lee [3], S.J. Park [3], T. O'Gorman [1], H. Lowe [1], M. Romanelli [1], M. Sertoli [1], M. Gemmel [1], J. Woods [1], H.V. Willett [1] & ST40 Team [1]

[1] Tokamak Energy Ltd., 173 Brook Drive Milton Park, Abingdon OX14 4SD, U.K.
[2] Rudolf Peierls Centre for Theoretical Physics, University of Oxford, Clarendon Laboratory, Parks Road, Oxford OX1 3PU, U.K.
[3] Department of Nuclear Engineering, Seoul National University, Seoul 08826, South Korea

E-mail: michail.anastopoulos@tokamakenergy.co.uk

## Abstract

In this paper, the turbulent transport properties of ST40 hot ion plasmas are examined and fully predictive time evolving modelling of a hot ion plasma pulse was performed. Understanding turbulent transport on spherical tokamaks (STs) is challenging due to their unique geometry characteristics. ST40 hot ion plasmas are typically unstable to ion scale Trapped Electron Modes (TEMs) and Ubiquitous Modes (UMs), driven from the kinetic response of trapped particles and passing ions, and electron scale Electron Temperature Gradient Modes (ETGs) at the edge of the plasma. A comparison between the linear unstable modes of the gyro-kinetic code GS2 and the gyro-fluid code TGLF showed that both models agree to a satisfactory level. However, some discrepancy was observed at the core of the plasma where a large fraction of beams ions exists, and electromagnetic effects are potentially important. Turbulent fluxes were also observed to be somewhat overpredicted with TGLF. The core heat ion transport is observed to be close to neoclassical levels due to turbulence suppression from high rotation and fast ion stabilisation, while the edge region is dominated by anomalous transport in both ions and electrons. As a result, enhanced energy confinement is observed in those plasmas driven by the reduced turbulent core region and the confined beam ions. Fully predictive simulations using the ASTRA transport solver coupled with SPIDER, NUBEAM, NCLASS and TGLF together with a novel reduced scrape of layer (SOL) model for the simulation of the last closed flux surface (LCFS) boundary conditions was attempted. Agreement in global quantities but also kinetic profiles between the predictive and interpretative modelling as well as experimental measurements was observed.

Keywords:

## 1. Introduction

Over the years, spherical tokamaks (STs) have demonstrated several beneficial characteristics, and provide an alternative path to a Fusion Pilot Plant (FPP) in comparison to conventional tokamaks (CTs). The small aspect ratio $A = R/r \leq 2$, where $r$ and $R$ are the minor and major radius respectively, typically leads to enhanced micro and macro stability properties as field lines spend more time in the good curvature region. STs can stably operate at higher elongation $\kappa$, which allows for higher bootstrap current, Greenwald density $n_{GW} = I_P/\pi^2 r$ and $\beta_N = \beta r B_T/I_P$, where $\beta = 2\mu_0 <p>/B^2$, $<p>$ is the volume averaged plasma pressure, $B_T$ is the toroidal magnetic field and $I_P$ is the plasma current. Several STs around the world have experimentally contributed to the creation of an ST database analysing the dependence of $\tau_E$ on engineering and dimensionless parameters. However, current ST devices do not operate in FPP relevant conditions. For this reason, a deep understanding of the physics, and particularly the turbulence and confinement characteristics, is required to effectively extrapolate and design an ST FPP using existing models.

Current state of the art nonlinear gyro-kinetic simulations have successfully shed light on the drive and saturation mechanisms for several of the observed turbulent phenomena in STs. Ion scale turbulence, related to ion temperature gradient (ITG) and kinetic ballooning mode (KBM), is strongly suppressed due to reduction of the bad curvature region along a field line and $d\beta/dr$ effects [1]. In addition, the large toroidal velocity shear [2][3] and fraction of beam ions [4][5][6] introduced by neutral beam heating is sufficient to suppress ion scale turbulence in the core. Electron scale electron temperature gradient (ETG) modes are also partially stabilised by the reduction of bad curvature region and $d\beta/dr$ effects. Although, the low aspect ratio can lead to trapped modes (TM) that can lead to significant transport [7][8][9]. Finally, due to higher $\beta$ STs are also characterised by electromagnetic turbulence produced mainly from unstable micro-tearing modes (MTM) [10][11]. The radial extent of occurring magnetic islands can be of the ion Larmor radius $\rho_i$ scale leading to large electron heat transport with velocity shear moderating those fluxes [12].

However, first principle nonlinear gyro-kinetic simulations are computationally intensive and can suffer from numerical instabilities. As a result, their use forbids extensive parameter scans for the optimisation and design of future devices. In recent years, machine learning surrogate models are being developed based on databases build from existing nonlinear gyro-kinetic simulations [13][14], but their applicability is still limited and at early stages. On the other hand, quasilinear transport models provide an alternative faster and promising route to perform such calculations. In CTs, such models have been successfully used and reproduced both qualitatively and quantitatively several plasma regimes [15][16][17]. Although, for STs such models have not been robustly predicting the experimentally measured plasma profiles [18][19][20]. There are several reasons why this is the case, but it can be narrowed down to the importance of shaping, electromagnetic and trapped particle physics that are typically approximated if not absent.

This work focuses on the use and applicability of the quasilinear model TGLF [21] as a reduced transport model to simulate ST40 hot ion plasmas [22]. Section(2) presents the ST40 hot ion plasma with details on the plasma parameters, geometry and kinetic profiles. In addition, a brief description on the interpretative and predictive modelling framework used is given. Section(3) focuses on the linear and nonlinear gyro-kinetic analysis of such plasmas using GS2 [23] and compares those results with the TGLF reduced transport model. Section(4) introduces the interpretative and predictive modelling workflow and presents fully predictive modelling using ASTRA [24] as a transport solver. Finally, Section(5) summarises the results.

## 2. ST40 hot ion plasma description

### 2.1 Integrated modelling of ST40 plasmas

Integrated modelling is used both as an interpretative analysis tool and as a predictive simulation tool. In this work, the plasma equilibrium and LCFS was calculated with EFIT using magnetic probes and matching the measured total plasma current. Electron density and temperature profiles were obtained from Thomson Scattering (TS) measurements [25]. Ion temperature and toroidal velocity were measured at the very core of the plasma from the charge exchange recombination spectroscopy (CXRS) diagnostic [26]. To obtain ion temperature profiles Bayesian analysis [27] is used to infer profiles that match the CXRS and X-ray crystal spectrometry (XRCS) measurements. Lastly, to obtain the ion density profile, interpretative modelling was employed. The transport code ASTRA is used, coupled with SPIDER [28] for the equilibrium reconstruction using the EFIT LCFS, and NUBEAM [29] for the calculation of beam ion density and power deposition. Through detailed interpretative modelling fully consistent kinetic profiles and equilibrium are obtained to perform the transport analysis. In fully predictive simulations transport models are required, and ASTRA is coupled to NCLASS [30] and TGLF. Finally, a reduced model for the SOL [31] is used to obtained boundary conditions (BCs) for density and temperature at the LCFS. Such information can be used to inform the location of the EFIT LCFS in comparison to TS measurements or used as BCs in fully predictive simulations. The experimental particle source is unknown, and it is set to be proportional to the edge particle flux and scaled to match the measured line-averaged density. Figure(2.1) gives a schematic of the modelling performed for ST40 plasmas.

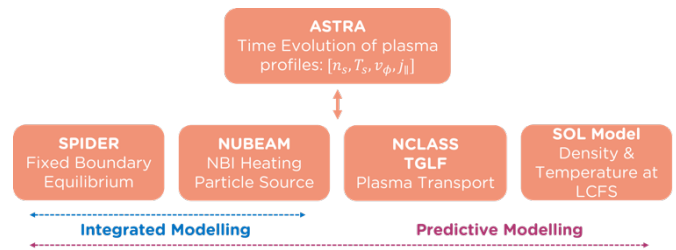

*Figure(2.1) Schematic of the integrated analysis tools used in interpretative and predictive simulations of ST40 plasmas.*

### 2.2 ST40 Hot ion plasma characteristics

ST40 is a high toroidal magnetic field spherical tokamak owned and operated by Tokamak Energy Ltd. One of its unique characteristics is that it can operate on a hot ion regime where $T_i > T_e$. In this paper, ST40 plasma pulse #11224 is examined in detail with respect to its transport characteristics.

The deuterium plasma parameters are given in Table(2.1). The toroidal field is $B_T$=1.8T and the plasma current $I_P$=550kA. The plasma is heated by two deuterium neutral beams which provide in total 1.8MW of heating power with significant toroidal rotational torque and core particle fuelling. The core toroidal velocity reaches $V_\varphi$=295km/s providing sufficient $\gamma_{E\times B}$ shear to suppress core turbulence. The aspect ratio is small with $A$=1.7 considering $R$=0.43m and $r$=0.25m. The elongation of the LCFS is typically moderate for those plasmas with $\kappa$=1.35 and triangularity is low with $\delta$=0.1. In this pulse, the core ion temperature was significantly higher than the electron with $T_i/T_e \sim 2.5$ corresponding to a core ion temperature of 5 keV.

| Hot Ion Plasma Pulse: #11224 | |
|---|---|
| $B_T$ [T] | 1.65 |
| $I_p$ [MA] | 0.55 |
| $R_0$ [m] | 0.43 |
| $\rho$ [m] | 0.25 |
| $\kappa$ | 1.35 |
| $\delta$ | 0.1 |
| $\bar{n}_e$ [m$^{-3}$] | $5.2 \cdot 10^{19}$ |
| $T_i/T_e$ | 2.5 |
| $\beta_N$ | 2.0 |
| $V_0$ [kms$^{-1}$] | 295.0 |
| $P_{NBI}$ [MW] | 1.8 |

*Table(2.1) Global plasma parameters and geometry for ST40 hot ion plasma pulse #1224.*

The time evolution of the measured plasma quantities can be seen from Figure(2.2). A time window exists between $t$=50-100ms where a slow varying state is reached that allows for reliable transport analysis to be performed. In addition, Figure(2.3) shows the electron kinetic profiles as measured from the TS diagnostic.

### 3. Transport analysis of ST40 hot ion plasma

In this section, the profiles and equilibrium depicted in Figure(2.3) are used to perform the gyro-kinetic analysis. Linear gyro-kinetic stability analysis is performed to understand the underlying turbulent modes that exist in ST40 hot ion plasmas using the GS2 code. Furthermore, nonlinear gyro-kinetic simulations examine the impact of beam ions and $\gamma_{E\times B}$ shear rate on the turbulent fluxes. Finally, the gyrokinetic results are compared with the reduced transport model of TGLF in terms of the linear and nonlinear characteristics of the model.

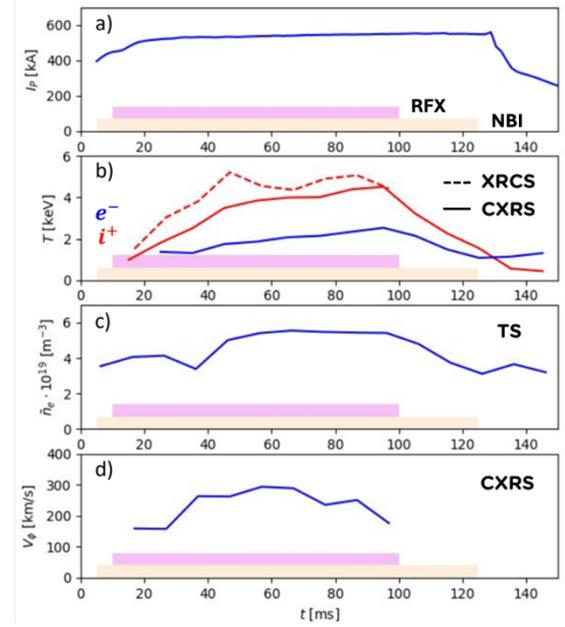

*Figure(2.2) Time trace of major global plasma parameters. a) Is the measured toroidal plasma current, b) the core temperature for ions and electrons as measured but CXRS and XRCS, c) the line average electron density as measured by TS and d) the core toroidal velocity as measured by the CXRS.*

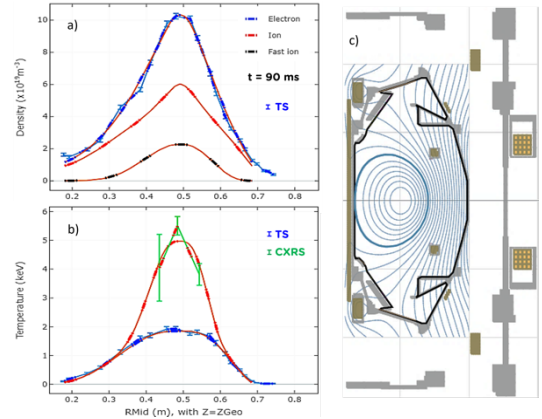

*Figure(2.3) Reconstruction of kinetic profiles and equilibrium geometry at t=90ms. a) Shows the electron, ion and beam ion density and electron density as measured by TS. b) Shows the electron and ion temperature and electron temperature as measured by TS and ion temperature as measured by CXRS. c) Depicts the EFIT equilibrium reconstruction of the flux surfaces from the magnetic measurements.*

**3.1 Gyro-kinetic properties of ST40 hot ion plasmas**

Figure(3.1) provides an overview of the linearly unstable modes that exist in a typical ST40 hot ion mode for various plasma radii. A zoo of micro-instabilities is observed similarly to previous ST40 hot ion mode gyro-kinetic analysis [32]. ETGs are observed mainly at the edge of the plasma and quickly get stabilised moving further in the core, as they are found stable for $\psi_N<0.5$ ($r_N<0.7$). Trapped modes, TEMs ($\omega<0$) and UMs ($\omega>0$), are unstable across the whole plasma radius. Moving from the mid-radius to the core region for $\psi_N<0.3$ ($r_N<0.55$) TEMs/UMs transition to ITGs, as it can be observed from the frequency and growth rate of the mode which becomes damped as $k_y\rho_i \to 1$. TEMs/UMs typically survive for $k_y\rho_i>1$. At the very core of the plasma for $\psi_N<0.1$ ($r_N<0.1$) KBMs are found unstable and are driven from the beam, which leads to core total pressure with $\beta\sim2\%$. Due to the large density gradient micro-tearing modes (MTMs) remain stable. Finally, as it can be observed from Figure(3.1) the estimated level of $\gamma_{E\times B}$ shear rate is sufficient to stabilise the core turbulence as it is larger than the growth rate of the instability. This is particularly important for stabilising core KBMs.

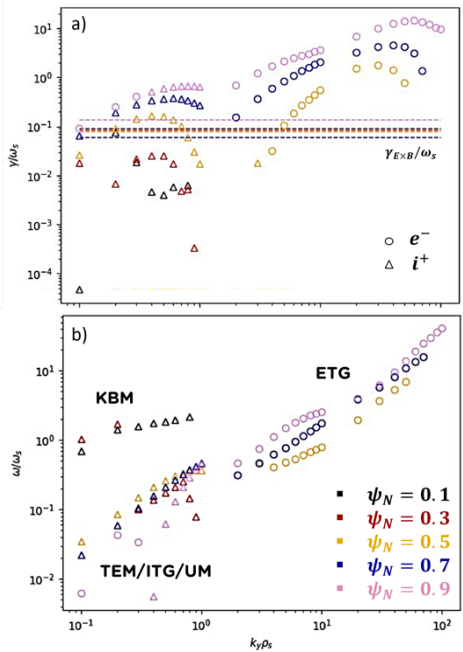

*Figure(3.1) a) show the normalised growth rates and b) normalised frequencies of the linearly unstable gyrokinetic mode for a wide range of $k_y\rho_i$ spanning all the way from the ion to electron scale. The dashed lines in a) as the estimated $\gamma_{ExB}$ shearing rate. The ▲ symbol indicates modes drifting in the ion direction ($\omega>0$), while the ● symbol indicates modes drifting in the electron direction ($\omega<0$).*

There are a couple of characteristics to consider in ST40 hot ion plasmas. The first is the peaked electron density and the second the large population of beam ions. Both of those features have profound implications for the underlying turbulence. The large density gradient, as well as the low aspect ratio, leads to large trapped particle drive and hence the observation of strongly unstable TEMs and UMs. The parametric dependence on inverse aspect ratio $\varepsilon$ and density gradient length scale $L_n$ is examined in Figure(3.2) and Figure(3.3). As it can be observed that reducing $\varepsilon$, which leads to the reduction of the trapped particle fraction, results in the stabilisation of the dominant instability. In addition, the decrease of $L_n$ results to further destabilisation of the dominant mode although $\eta$ has decreased. This is indicative of trapped modes which are driven from the density gradient. In addition, TEMs survive at larger $k_y\rho_i$ until they continuously transition to UMs. Although, decreasing the temperature gradient length $L_T$ (increasing $\eta$) is also a drive for UMs as can be observed from Figure(3.4). For such plasmas a strong coupling between trapped electrons and passing ions seems to exist.

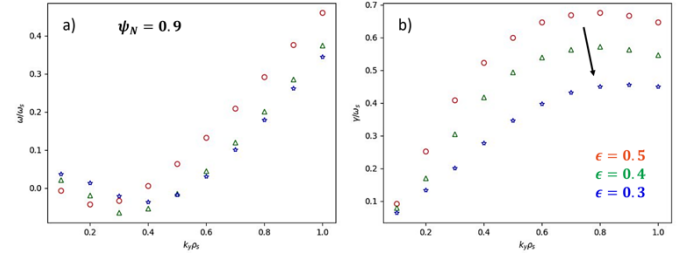

*Figure(3.2) a) Normalised frequency $\omega/\omega_s$ and b) normalised growth rate $\gamma/\omega_s$ as a function of the binormal wavenumber $k_y\rho_s$ for the dominant linear instability at $\psi_N=0.9$ with varying aspect ratio $\varepsilon$.*

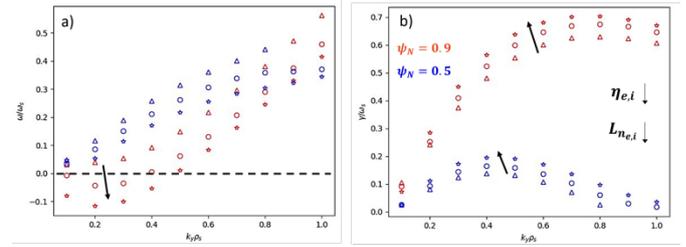

*Figure(3.3) a) Normalised frequency $\omega/\omega_s$ and b) normalised growth rate $\gamma/\omega_s$ as a function of the binormal wavenumber $k_y\rho_s$ for the dominant linear instability at $\psi_N=0.9$ and $\psi_N=0.5$ with varying the density gradient length $L_n$.*

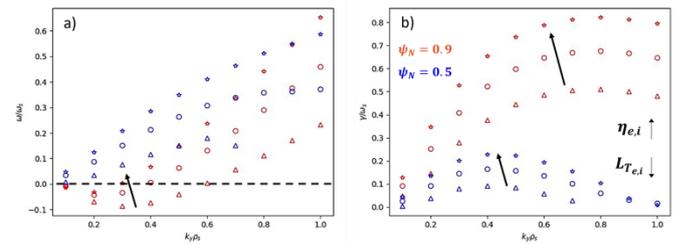

*Figure(3.4) a) Normalised frequency $\omega/\omega_s$ and b) normalised growth rate $\gamma/\omega_s$ as a function of the binormal wavenumber $k_y\rho_s$ for the dominant linear instability at $\psi_N=0.9$ and $\psi_N=0.5$ with varying the temperature gradient length $L_T$.*

Secondly, the beam ions have a stabilising effect on turbulence as it can be observed from Figure(3.5). There are two main mechanisms for this stabilisation. Due to their high energy they lead to an increase in $\beta$ and therefore $d\beta/dr$. In addition, they decrease the density gradient of the main ions through dilution. A similar dilution effect is observed from including impurities. However, because the beam allows the plasma core to reach higher $\beta$, KBMs can be destabilised. As it will be shown, with sufficient $\gamma_{ExB}$ shear rate those modes can be stabilised.

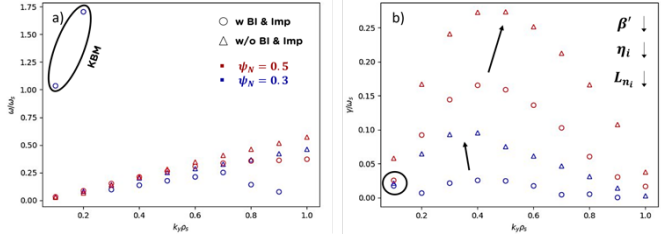

*Figure(3.5) a) Normalised frequency $\omega/\omega_s$ and b) normalised growth rate $\gamma/\omega_s$ as a function of the binormal wavenumber $k_y\rho_s$ for the dominant linear instability at $\psi_N=0.5$ and $\psi_N=0.3$ with and without beam ions and impurities.*

The impact of fast ions and $\gamma_{E\times B}$ shear rate is examined on the nonlinear gyro-kinetic turbulent flux at $\psi_N=0.5$ using GS2. It is observed that $\gamma_{E\times B}$ shear rate is a key mechanism for turbulence suppression. The velocity shear rate is computed through interpretative modelling. The torque as computed from NUBEAM is used as a source to evolve a diffusion equation for the toroidal velocity $V_\varphi$. The momentum diffusion coefficient is considered to have the same functional form as the ion diffusion coefficient and scaled to match the measured core toroidal velocity in time, i.e. $\chi_\varphi=c\chi_i$ with $c\sim0.7$. In this way an estimate of $\gamma_{E\times B}$ shear rate can be obtained. As it can be observed from Figure(3.6) 15% increase of $\gamma_{E\times B}$ can lead to the complete suppression of heat flux. Moreover, it is found that beam ions do have a significant stabilising effect, reducing the turbulent heat flux by 30% as it can be seen in Figure(3.7). This is consistent with the linear stability where the inclusion of beam ions led to reduced growth rates. As a result, the combination of beam ions and strong rotation leads to the reduction of transport allowing the establishment of large core temperature. It needs to be noted that the transport of impurities and beam ions is rather in comparison to the main ion and electron species.

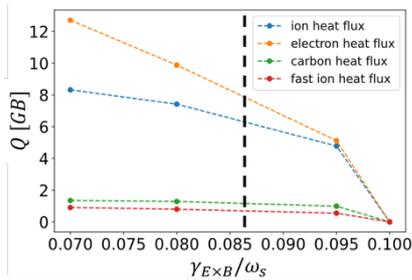

*Figure(3.6) Nonlinear gyro-kinetic heat flux Q [GB] at $\psi_N=0.5$ as a function of $\gamma_{ExB}$ shear rate. The dashed line indicates the estimated equilibrium value considering the rotation and pressure gradient profiles.*

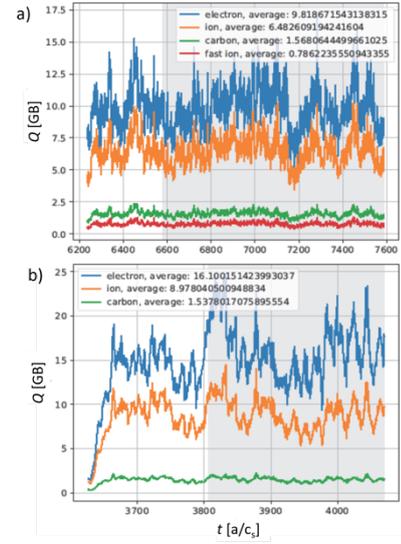

*Figure(3.7) Comparison of the nonlinear gyro-kinetic heat flux Q [GB] at $\psi_N=0.5$ a) with beam ions and b) without beam ions.*

### 3.2 Comparison of gyro-kinetics with the TGLF model

Reduced transport models have been proven an extremely useful tool in predictive modelling of CTs due to their supreme balance between accuracy and speed. However, their application on STs has been limited due to the lack of robust predictive capabilities. TGLF is the most complete reduced transport model to date. Its applicability in STs has remained questionable due to its partial success in MAST [19] and its lack of reproducing NSTX [20] high performance plasma pulses.

ST turbulence related to ITG is typically characterised by suppressed ion fluxes from the high rotation as well as the enhanced electromagnetic effects. Therefore, the calibration of the saturation mechanism and amplitude that TGLF is based on can deviate in STs. In addition, nonlinear gyro-kinetic simulations of MTM turbulence for MAST and NSTX plasmas result in significant electron heat transport. Such modes are not accurately captured by the TGLF linear solver. In NSTX, coupled TEMs/KBMs are also observed to drive transport, but due to the lack of $\delta B_\parallel$ effects transport will be underestimated.

In ST40 plasmas, the applicability of the TGLF model, and particularly the SAT2 model [33][34], is examined. ST40 due to the high $B_T$ is expected to demonstrate reduced electromagnetic turbulence (apart from the very core where $\beta\sim O(1)\%$) as KBMs are stabilised by rotational shear and MTMs are stable. In addition, pedestal formation is not observed in hot ion plasmas and so TGLF could be applicable for a full radius predictive simulation. A large question that remains is the accuracy of the rotational physics and the effect of beam ions on turbulence stabilisation.

To begin with, it is important to understand if TGLF linear solver captures the correct physical instabilities. Figure(3.8) shows a comparison of the frequency and growth rate for a wide range of $k_y\rho_i$ spanning from the sub-ion scale all the way to the electron scale. As it can be observed, the two models agree well with its other. It should be noted that this agreement becomes worse at the core of the plasma where $\beta$ and the fraction of beam ions significantly increases. TGLF fails to capture the unstable KBMs observed with GS2, as $\delta B_\parallel$ is excluded from the TGLF model. Although, excluding the KBMs, the modes at marginal stability for $\psi_N<0.3$ are expected to deviate between the two codes. Overall, a good agreement is observed for the strongly driven modes for $\psi_N>0.5$ and TGLF correctly captures that the dominant drive comes from trapped particle effects.

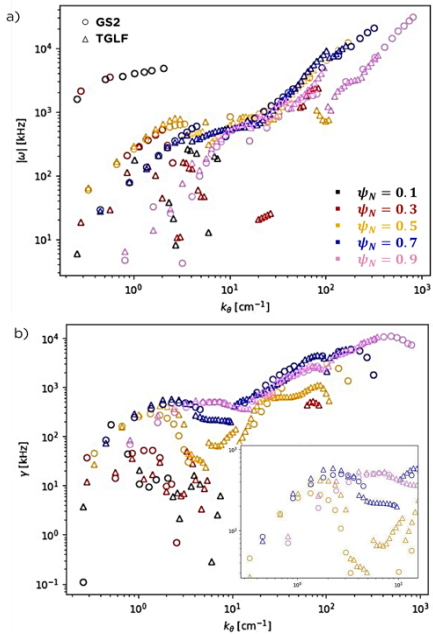

*Figure(3.8) Comparison of GS2 ● and TGLF ▲ a) normalised frequency $\omega/\omega_s$ and b) normalised growth rate $\gamma/\omega_s$ as a function of the binormal wavenumber $k_y\rho_s$ for various flux surfaces.*

Table(3.1) shows the comparison of the nonlinear fluxes between GS2 and TGLF for the equilibrium $\gamma_{E\times B}/\omega_s\sim 0.086$ shear rate. Although, a qualitative agreement is observed, GS2 predicts somewhat lower heat fluxes by 30% in comparison to TGLF which can have a significant impact on predictive simulations.

| Code / [GB] | $Q_e$ | $Q_i$ | $Q_c$ | $Q_{beam}$ |
|---|---|---|---|---|
| GS2 | 7.8 | 6.3 | 1.2 | 0.7 |
| TGLF | 10.8 | 9.6 | 1.6 | 0.35 |

*Table(3.1) Comparison of GS2 and TGLF heat flux Q [GB] at at $\psi_N=0.5$ considering the estimated equilibrium $\gamma_{E\times B}/\omega_s\sim 0.086$ shear rate.*

## 3.3 Comparison of the TGLF model with the interpretative heat fluxes

Although TGLF and GS2 were in good agreement, it needs to be noted that the nonlinear fluxes are found to be several orders of magnitude larger than what the interpretative analysis would suggest. This is attributed to the stiff transport regime and the uncertainty in profile gradients, mainly the ion kinetic profiles and the toroidal velocity profile. For those reasons a parametric scan is performed in $\gamma_{E\times B}$ to examine the sensitivity to rotational effects. As it can be observed from Figure(3.9), $\gamma_{E\times B}$ is responsible for the suppression of turbulence at the core of the plasma. However, even doubling $\gamma_{E\times B}$ still leads to significantly higher heat fluxes. In addition, the TGLF rotational model seems to be less sensitive to $\gamma_{E\times B}$ in comparison to the nonlinear simulations done with GS2 for this case.

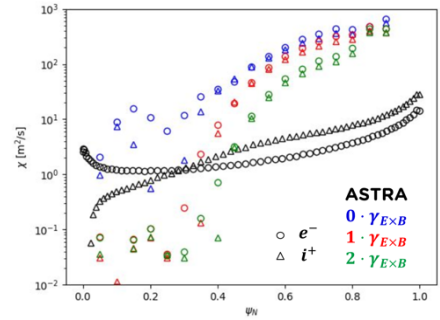

*Figure(3.9) Comparison of TGLF electron ● and ion ▲ heat flux Q [GB] as a function of normalised poloidal flux $\psi_N$ with varying $\gamma_{E\times B}/\omega_s$ shear rate.*

Moreover, the heat fluxes as computed from TGLF are highly driven from trapped particle effects, which is consistent with the linear gyro-kinetic analysis where trapped particles are a strong drive of turbulence. Figure(3.10) shows a scan of $\theta_{trapped}$, a parameters that scales the bounce location along the field line and therefore affects the amplitude of bounce averaging integrals. Reducing $\theta_{trapped}$ below $\theta_{trapped}<0.5$ results in stabilisation of the linear modes and strong deviation from gyro-kinetic linear spectrum.

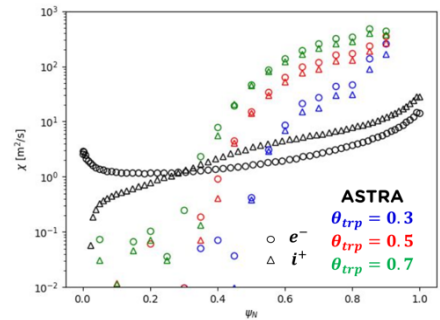

*Figure(3.10) Comparison of TGLF electron ● and ion ▲ heat flux Q [GB] as a function of normalised poloidal flux $\psi_N$ with varying $\theta_{trapped}$.*

Finally, the beam ions and impurities were excluded from the TGLF simulation to assess their impact. It can be observed that they play a stabilising role reducing the heat fluxes as it can be seen from Figure(3.11), again in qualitative agreement with the nonlinear GS2 simulations.

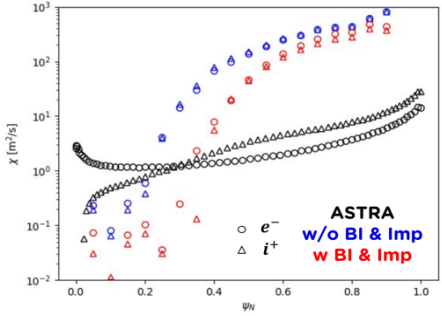

*Figure(3.11) Comparison of TGLF electron ● and ion ▲ heat flux Q [GB] as a function of normalised poloidal flux $\psi_N$ without (red) and with (blue) beam ions and impurities.*

## 4. Predictive transport simulation of ST40 hot ion plasmas

The above analysis revealed that TGLF is in satisfactory agreement with gyro-kinetic simulations. Although much larger transport was observed, this can be attributed to the stiffness of turbulence on various input parameters, like the assumed ion, impurity and toroidal rotation profiles. In this section, fully predictive modelling is employed to assess the sensitivity of the system to such assumptions. The self-consistent evolution of the profiles with the equilibrium and the transport will allow to properly assess the applicability of the TGLF model as a potential transport model for ST40 hot ion plasmas.

The procedure used to perform these simulations is depicted in Figure(2.1). ASTRA is used as a transport solver to evolve electron and main ion density $n$ and temperature $T$, toroidal plasma current density $j_\varphi$ and toroidal velocity $V_\varphi$. ASTRA is coupled with SPIDER that is used as a fixed boundary Grad-Safranov solver to calculate the equilibrium flux surfaces for given plasma profiles. The EFIT boundary is used as a reference for the plasma edge in time. Heating, particle and momentum sources are obtained from NUBEAM that is set to the two beam launcher configurations of ST40. Additional particle source coming from the wall/vacuum area is also considered and such a module exist already within ASTRA. The neoclassical and anomalous transport coefficients of particle $D$ and heat $\chi$ diffusion are obtained using NCLASS and TGLF. The convective velocity due to neoclassical or anomalous effects is also included in the simulation. For the momentum diffusion it is assumed that $\chi_\varphi=0.7\chi_i$, and this coefficient is chosen such that core rotation is in good agreement with experimental measurements. It needs to be stressed that the goal of evolving $V_\varphi$ is to provide an estimate of $\gamma_{E\times B}$ rather than a consistent evolution of the rotation profile. Finally, a novel addition to the simulation is the introduction of self-consistent edge boundary conditions for the electron density and temperature as well as ion temperature. The model is based on an extended two-point SOL model that includes power and particle balance. More details can be found in Ref[31].

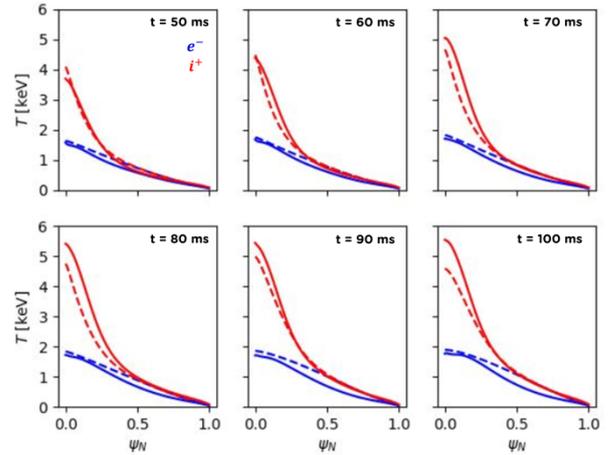

*Figure(4.2) Comparison of electron and ion interpretative (dashed line) and predictive (solid line) temperature as a function of normalised poloidal flux $\psi_N$ for a number of different time slices t.*

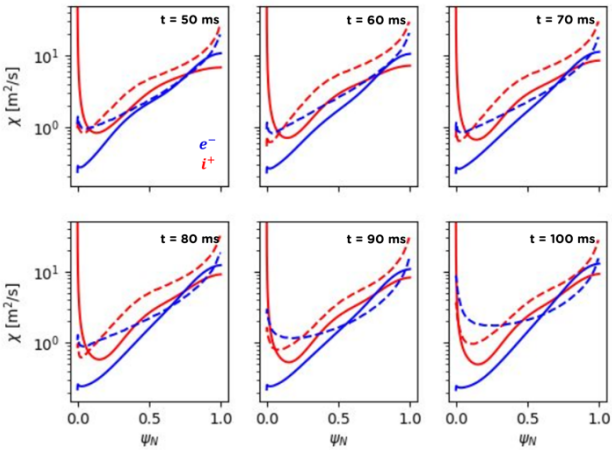

*Figure(4.1) Comparison of electron and ion interpretative (dashed line) and predictive (solid line) heat diffusion coefficient $\chi$ as a function of normalised poloidal flux $\psi_N$ for a number of different time slices t.*

The fully predictive simulation resulted in a good agreement between experimental profiles and interpretative profiles. This reveals the large sensitivity of the anomalous transport to profile gradients, as the interpretative profiles led to very large transport. As it can be observed from Figure(4.1) the predicted ion heat diffusion is in good agreement with the interpretative analysis, although a smaller value is obtained. The main difference is the higher electron transport towards the edge region and lower transport towards the core. This leads to narrower electron temperature profile in comparison to the measured TS $T_e$ with lower temperature gradients at the mid-edge region. This resulted in significant reduction in the predicted transport coefficients for both ions and electrons in comparison to the TGLF result associated with the interpretative profiles. Although, the smaller ion heat diffusion the predicted ion profiles are in good agreement with the interpretative ones. Figure(4.2) shows the comparison of interpretative and predictive temperature.

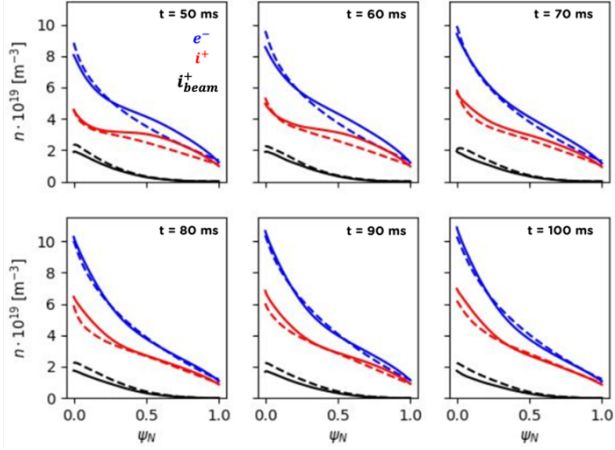

*Figure(4.3) Comparison of electron and ion interpretative (dashed line) and predictive (solid line) density as a function of normalised poloidal flux $\psi_N$ for a number of different time slices t.*

Moreover, the comparison of electron density $n_e$ between the predictive simulation and TS measurements, resulted in good agreement, as can be seen from Figure(4.3). A key ingredient for this agreement is the introduction of adaptive wall neutral source to keep the electron line-average density the same as the experimental value. The level of particle flux to match the measured line-averaged density corresponds to ~$10^{16}$ particles in the vacuum chamber. However, experimental measurements do not exist for the neutral content in the vacuum region to assess the validity of these values, and therefore the level of particle transport as predicted by TGLF cannot be easily verified.

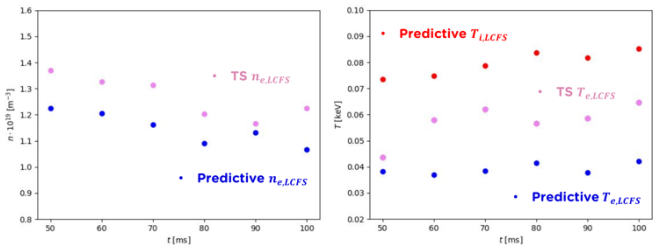

*Figure(4.4) Comparison of a) electron LCFS density and b) electron and ion LCFS temperature between the predictive simulation using the SOL model and the measured TS values.*

The SOL model requires as inputs the heat and particle flux, which are obtained from TGLF, and an estimate of the decay length $\lambda_q$ and connection length $L$ of the SOL region. To estimate those values, we assume that $\lambda_q \sim 2\lambda_q^{Eich}$ [35], considering the edge of a hot ion mode is like an L-mode plasmas, and $L \sim 2\pi qR$ at the plasma edge. As it can be seen from Figure(4.4) a good agreement is observed for the simulated electron density at the edge and the TS measurement, with a difference of 15%. For the electron temperature, it seems that $T_e$ is underestimated in comparison to the TS measurement by 35%, with the ion temperature to be almost double that of the electrons, i.e. $T_{i,LCFS} \sim 2T_{e,LCFS}$. It is not unlikely that this discrepancy in $T_{e,LCFS}$ exists due to errors between the true location of the LCFS between EFIT and TS. The assumption with respect to $\lambda_q$ and $L$ or the TGLF fluxes and impurity concertation do contribute to this discrepancy as well. It needs to be stressed though that the agreement is good.

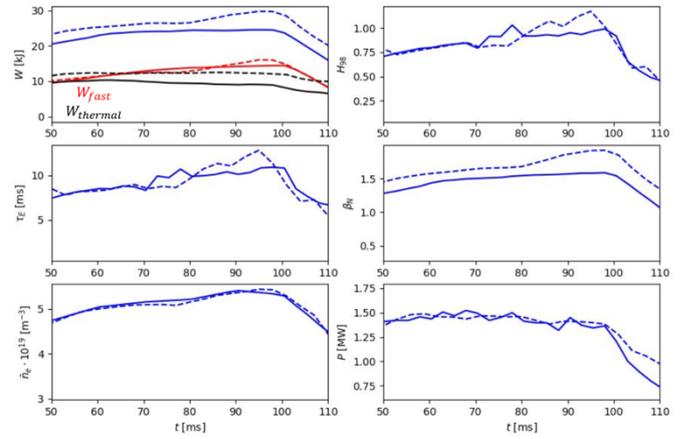

*Figure(4.5) Comparison of various global plasma parameters between the interpretative (dashed line) and predictive (solid line) simulations.*

Finally, Figure(4.5) shows the comparison of global parameter, in terms of stored energy $W$ [kJ], $H_{98}$, energy confinement time $\tau_E$ [ms], $\beta_N$, line averaged density $n_e$ and total absorbed power $P$ [MW], between the predictive and interpretative simulations. There is a good agreement between the two simulations, with the main difference being the underprediction in the stored energy by 20%, and therefore $\beta_N$, due to more narrow temperature profile for the electrons. This is depicted in the thermal component of the stored energy. Overall, the predictive simulation reproduced the interpretative global quantities.

## 5. Summary

To sum up, this work focused on examining the transport properties of ST40 hot ion plasmas using integrated modelling tools for interpretative and predictive simulations. To begin with, linear gyro-kinetic analysis was employed to understand the underlying unstable turbulent modes that exist in a typical ST40 hot ion plasma. It was observed that trapped particle effects are the main mechanism for driving turbulence in those plasmas, due to the small aspect ratio and the peaked electron density profile. Using nonlinear gyro-kinetic simulations, the beam ions and impurities are observed to have a stabilising effect leading to reduced transport, but at the core of the plasma rotational $E \times B$ shear seems to be the main mechanism through which transport is suppressed, and large $T_i$ is reached.

The comparison of the gyro-kinetic analysis with GS2 to the TGLF SAT2 model resulted in a good overall agreement, with TGLF qualitatively capturing the correct micro-instabilities, but turbulent fluxes were ~30% larger. Using the interpretative profiles, the TGLF fluxes showed a huge discrepancy with the interpretative power balance. TGLF fluxes were order of magnitude larger which suggested that ST40 hot ion plasmas are found in a stiff transport regime. Accurate measurements for the ion temperature and toroidal rotation profiles are necessary to verify TGLF but also the interpretative levels of transport. The examination of the interpretative profiles alone can be therefore misleading.

Moreover, performing fully predictive transport simulations with TGLF as the anomalous transport model resulted in good agreement between the experimental electron density and temperature as measured by the TS diagnostic system. This fact reveals the sensitivity of the reduced transport model to profile gradients and the stiff nature of the turbulent fluxes. However, the electron transport at mid-edge region seems to be overestimated by TGLF. Future work will focus on validating reduced quasi-linear models at the plasma edge as it is often a region where multi-scale and non-local effects can manifest. In addition, the reduced SOL model resulted in good agreement with the measured edge density and temperature conditions. Several key global parameters, where also reproduced by the predictive simulations.

Finally, there are a few key areas that require further improvement to qualitatively assess the validity of both TGLF and the reduced SOL model. Better experimental quantification of the impurity concentration and impurity density profiles are required. The impurities can have an important impact on diffusive turbulent flux as well as convective particle velocity and electron density peaking. Particle fuelling from the vacuum region is also a very important aspect of predictive modelling due to its major implications for the electron density. In addition, the SOL model depends on $\lambda_q$ and $L$. In ST40 hot ion plasmas such measurements are not available. Although from recent diverted configurations, $\lambda_q$ was measured and future work will focus on assessing the accuracy and applicability of the reduced SOL model on those cases.

**Acknowledgements**

The authors are greatful for the hard work and dedication of the ST40 operations and diagnostics team.